# FORS-EMG: A Novel sEMG Dataset for Hand Gesture Recognition Across Multiple Forearm Orientations

Umme Rumman, Arifa Ferdousi, Bipin Saha, Md. Sazzad Hossain, Md. Johirul Islam*, Shamim Ahmad, *Member, IEEE*, Mamun Bin Ibne Reaz*, *Senior Member, IEEE*, Md. Rezaul Islam

*Abstract*— Surface electromyography (sEMG) signals hold significant potential for gesture recognition and robust prosthetic hand development. However, sEMG signals are affected by various physiological and dynamic factors, including forearm orientation, electrode displacement, and limb position. Most existing sEMG datasets lack these dynamic considerations. This study introduces a novel multichannel sEMG dataset to evaluate commonly used hand gestures across three distinct forearm orientations. The dataset was collected from nineteen able-bodied subjects performing twelve hand gestures in three forearm orientations—supination, rest, and pronation. Eight MFI EMG electrodes were strategically placed at the elbow and mid-forearm to record high-quality EMG signals. Signal quality was validated through Signal-to-Noise Ratio (SNR) and Signal-to-Motion artifact Ratio (SMR) metrics. Hand gesture classification performance across forearm orientations was evaluated using machine learning classifiers, including LDA, SVM, and KNN, alongside five feature extraction methods: TDD, TSD, FTDD, AR-RMS, and SNTDF. Furthermore, deep learning models such as 1D CNN, RNN, LSTM, and hybrid architectures were employed for a comprehensive analysis. Notably, the LDA classifier achieved the highest F1 score of 88.58% with the SNTDF feature set when trained on hand gesture data of resting and tested across gesture data of all orientations. The promising results from extensive analyses underscore the proposed dataset's potential as a benchmark for advancing gesture recognition technologies, clinical sEMG research, and human-computer interaction applications. The dataset is publicly available in MATLAB format. Access the dataset here.

## I. INTRODUCTION

IN the human body, the human hand is one of the sophisticated and versatile organs that enable individuals to perform daily activities [1], [2]. It enables people to create, communicate, and use tools to perform various movements within their environments [3]. However, amputees encounter significant challenges due to the loss of a hand or varying levels of amputation, often leading to permanent impairments that affect their ability to work, maintain independence, and enjoy a good quality of life [4]. Hand gesture recognition (HGR) systems are gaining popularity because they enable integrated and efficient communication between humans and machines. These systems, along with prosthetic hands, are designed to enhance or restore the appearance and functionality of the human hand, enabling users to perform routine tasks. These devices represent a way for prosthesis users to regain lost hand functionality. Therefore, a reliable and effective gesture recognition system is essential for developing a robust human-machine interface [5].

Surface electromyography (sEMG) signals are increasingly important as biological metrics, with applications in human-machine interfacing, prosthetic technology, and rehabilitation devices. sEMG measures electromyographic potential, offering valuable insights into sensory activity and detailed movement control for prosthetic limbs [6]. Both non-invasive (surface electrodes) and invasive (needle electrodes) techniques can be used to record electromyogram (EMG) signals. Currently, sEMG is the most widely used non-invasive method in gesture recognition research for assessing muscle activity [7], [8], [9]. In smart prosthetics, sEMG signals are processed using regression and pattern recognition methods, involving preprocessing, feature extraction, dimension reduction, and machine learning for hand gesture classification [10], [11]. Advanced myoelectric prosthetics like the Vincent Hand Evolution 2, Otto Bock Michelangelo, and Touch Bionics i-limb Quantum enable complex movements but face challenges in control robustness and speed due to the noisy and variable nature of sEMG signals [12],[13], [14]. sEMG signals naturally become unstable and are influenced by various physiological

This work was financially supported by the Ministry of Science and Technology, Bangladesh under reference grant code: SRG-232388, and Rajshahi University of Engineering & Technology (RUET) under grant no DRE/7/RUET/574(58)/PRO/2023-24/50.

*Corresponding Authors: Mamun Bin Ibne Reaz and Md. Johirul Islam
Umme Rumman is with the Department of Computer Science and Engineering, Varendra University, Rajshahi, Bangladesh (email: chaiti@vu.edu.bd).
Arifa Ferdousi is with the Department of Electrical, Electronic, and Systems Engineering at Universiti Kebangsaan Malaysia, Bangi, Selangor, Malaysia (e-mail: afpris@gmail.com).
Md. Johirul Islam is with the Department of Physics, Rajshahi University of Engineering and Technology, Rajshahi-6204, Bangladesh (e-mail: johirul@phy.ruet.ac.bd).
Bipin Saha, Md. Sazzad Hossain are with the Department of Electrical and Electronic Engineering, University of Rajshahi, Rajshahi-6205, Bangladesh (e-mail: bipinsaha.bd@gmail.com, shsajal111@gmail.com).
Shamim Ahmad is with the Department of Computer Science and Engineering, University of Rajshahi, Rajshahi-6205, Bangladesh (e-mail: shamim_cst@ru.ac.bd).
Mamun Bin Ibne Reaz is with the Department of Electrical and Electronic Engineering, Independent University, Dhaka, Bangladesh (e-mail: mamun.reaz@iub.edu.bd).
Md. Rezaul Islam is with the Department of Electrical and Electronic Engineering, University of Rajshahi, Rajshahi-6205, Bangladesh (e-mail: rima@ru.ac.bd).

and environmental factors, such as fatigue, sweat, forearm orientation, muscle contraction force, electrode shift, multi-day variations, and limb position. These variations can impact the amplitude, frequency distribution, and features or shape of the signal, thereby affecting the robustness of the control methodologies of prosthetic hands [13], [15], [16], [17].

A substantial amount of research has focused on investigating the influences of these dynamic factors on sEMG pattern recognition (sEMG-PR), yielding optimistic outcomes in classification (typically exceeding 90%) and real-time performance [18], [19]. One study explored the impact of variable arm positions on the proficiency of hand gesture classifiers at distinct static postures and dynamic forearm motions. The outcomes highlighted that the classifier achieved better accuracy of classification for the dynamic forearm movement than the static position [20]. Another study examined the impact of electrode shifting on the effectiveness of pattern recognition. They found that such displacement of the electrode adversely impacted the classification performance. In particular, the displacement of the electrode by up to 1 cm from its initial training position dropped the accuracy of the classification from about 90 % to 60 %. However, they suggested that if the system can be trained with plausible displacement locations, this adverse impact might be reduced [21]. Research on the variation of joint angle and muscle contraction level revealed that both factors affect sEMG signal parameters, estimating mean frequency and conduction velocity. This study showed that forearm orientation changes had a greater impact on sEMG signals compared to muscle contraction force levels [22]. Rajapriya et al. [37] investigated the effects of forearm orientation and contraction force on sEMG signals, introducing wavelet bispectrum-based features to improve classification accuracy. Similarly, Islam et al. [26] proposed a feature selection method to identify the minimal dimensional features needed to enhance sEMG-PR performance under these variations. Several studies have also investigated the impact of electrode shifting, muscle contraction level variations, and fatigue on sEMG pattern recognition performance. The research found that electrode movement and variations in muscle contraction levels significantly affect the sEMG-PR performance more than muscle fatigue [23]. A recent study highlighted that hand gesture recognition performance is significantly influenced by forearm posture, position, and orientation during daily tasks. However, this impact can be minimized by utilizing accelerometers, incorporating a wide range of features, and training the pattern recognition system with data from different real-time positions, forearm postures, and orientations [24], [25], [26].

Table I presents several publicly available datasets for hand gesture recognition. The Non-Invasive Adaptive Hand Prosthetics (NinaPro) is the largest multimodal database. A total of 10 Ninapro datasets contain more than 180 data acquisitions on electromyography, kinematic, inertial, clinical, neurocognitive, and eye-hand coordination from intact subjects and transradial hand amputees. NinaPro DB1 [27] and DB4 [28] consist of 52 gestures performed by 27 (DB1) and 10 (DB4) subjects recorded with 10 and 12 sparsely located electrodes, respectively. In addition, NinaPro DB7 [29] includes 40 gestures performed by 20 subjects, recorded with 12 sparsely spaced electrodes. Although these datasets are widely used in the research community, they didn't account for dynamic factors affecting sEMG signals, such as forearm orientation, different electrode localization sites, various muscle contraction force levels, etc. CapgMyo (DB-a)[30] recorded signals from eight gestures using 128 electrodes on 18 subjects. However, it lacks a sufficient number of hand gestures commonly performed daily. Forearm orientation and muscle contraction force level were the two crucial dynamic factors considered in the Rami-Khushaba [31] dataset; however, it included only a limited number of gestures and subjects. A moderately high number of repeats of some complex daily living gestures were taken into account by EMAHA-DB [32], but no influential factors were taken into account when the data was being recorded. To build more robust models that can adapt to different real-world conditions, it's crucial to collect data that reflects the variability introduced by forearm orientation, electrode localization, contraction force levels, and other dynamic factors.

In this study, we measured muscle activity related to everyday activities and collected novel multichannel sEMG signals from the Bangladeshi population. The sEMG signal was recorded using eight channels for twelve active daily living hand gestures. Compared to existing sEMG datasets, the key features of the proposed dataset are: -

1) **Forearm orientations:** Three forearm orientations (pronation, rest, and supination) were considered during gesture performance, as they are key dynamic factors that influence sEMG signal parameters.
2) **Electrode placement**: Electrodes were placed at two distinct locations on the right arm (elbow and the middle of the forearm), which facilitates the analysis of sEMG signals from different forearm locations, particularly for patients with wrist disarticulation and transradial amputations.
3) **Reduced complexity**: Instead of using high-density electrodes, the sEMG signal was recorded using 8 electrodes, reducing the complexity of the pattern recognition system.
4) **Diversity**: Compared to the existing dataset on forearm orientations, the proposed dataset includes a larger number of subjects, a greater variety of gestures, and a sufficient number of gesture repetitions.
5) **Quality assessment**: The quality of the recorded sEMG signals was assessed using two signal quality parameters. Additionally, the performance of the dataset was analyzed through comprehensive feature extraction using five well-established methods. Cutting-edge machine learning and deep learning models, including LDA, SVM, KNN, CNN, RNN, LSTM, and hybrid models, were employed to classify different hand gestures.

The paper is organized as follows: Section II details the methodology for recording, protocol, and analysis of the sEMG signals. Section III presents the results of various analyses. The implications of these results, along with future directions, are discussed in Section IV. Finally, Section V concludes the paper.

TABLE I
COMPARISON WITH BENCHMARKED DATASET ACCORDING TO BASIC DATA CHARACTERISTICS

| Reference | Dataset Name | Action Type | Number of Channels | Number of Subjects | Number of Gestures | Dynamic factor | Number of Trials | Gesture Duration (s) | Classification Model | Total Signals |
|---|---|---|---|---|---|---|---|---|---|---|
| Atzori et al. [27] | NinaPro DB1 | Hand and wrist, gestures, gripping objects | 10 | 27 | 52 | N/A | 10 | Gesture: 5s Rest: 3s | Least-Squares SVM | 14040 |
| Pizzolato et al. [28] | NinaPro DB4 | Hand and wrist, gestures, gripping objects | 12 | 10 | 52 | N/A | 6 | Gesture:5s Rest: 3s | RF, SVM | 3120 |
| Krasoulis et al. [29] | NinaPro DB7 | Hand and wrist, gestures, gripping objects | 12 | 20 | 40 | N/A | 6 | Gestures: 5s Rest: 5s | LDA | 4800 |
| Du et al. [30] | CapgMyo (DB-a) | Isotonic and isometric hand gestures | 128 | 18 | 8 | N/A | 10 | Gesture:3s Rest: 7s | ConvNet | 1440 |
| Khushaba et al. [31] | Rami-Khushaba | Hand gestures | 8 | 13 | 6 | Forearm orientation, muscle contraction level | 3 | Gesture:5s Rest: 10s | SVM, LDA | 2106 |
| Karnam et al. [32] | EMAHA-DB | Daily activities: Gripping, holding, writing, and drawing | 5 | 25 | 22 | N/A | 10 | Gesture:5-15s Rest: 3 to 5s | RF, LDA, FKNN, sKNN, SVM3 and CNN+LSTM | 5500 |
| **Proposed** | **FORS-EMG** | **Finger and wrist gestures** | **8** | **19** | **12** | **Forearm orientation, elbow to mid-forearm electrode position** | **5** | **Gestures: 8s Rest: 3 to 5s** | **LDA, SVM, KNN, 1D CNN, RNN, and LSTM** | **3420** |

## II. METHODOLOGY

### A. Subjects

In this experiment, nineteen healthy subjects participated in the sEMG data collection procedure. Among the participants sixteen were males and three were females, with ages ranging from 25 to 40 years. Before data collection, each subject signed a participation and data publication consent. Additionally, they filled out a form giving information about their age, height, and medical history. All subjects were confirmed to be free from major diseases, and none reported any conditions that might interfere with the sEMG data collection process. Ethical approval was issued by the Dean of Applied Science and Humanities at Rajshahi University of Engineering & Technology for this study.

### B. Experimental Protocol

During data recording, subjects were asked to perform twelve active daily living hand gestures as shown in Fig. 1(c). The gestures included three single-finger, five multi-fingers, and four wrist gestures. The twelve active gestures include the following: thumb up (TU), index (IDX), right angle (RA), peace (PCE), index little (IL), thumb little (TL), hand close (HC), hand open (HO), wrist extension (WE), wrist flexion (WF), ulnar deviation (UD), and radial deviation (RD). Before data collection, participants were provided with a detailed visual demonstration of the acquisition process, which included step-by-step instructions on how to perform each hand gesture in the required sequence. Participants were allowed to practice the gestures multiple times to become familiar with the experimental protocol and ensure consistency in execution. After the participants were acquainted with the experimental protocol, the session of the respective subjects for sEMG data collection was started. In addition, to enhance the signal quality, a conductive gel was initially applied to the sensor and skin underlying the targeted muscles. All sEMG signals of each subject were collected within the same session. An elastic band was placed over the electrodes to ensure that their positions remained consistent throughout the recording process. Before further analysis, the quality of the collected sEMG signal was visually inspected by sEMG field expertise. In this acquisition, three forearm orientations (Fig. 1(d)) were considered, supination, rest, and pronation. At each forearm orientation, the subject performed all twelve gestures of 8 s duration and repeated five times for each gesture from a relaxed position. Instruction was given to subjects continuously before performing successive gestures with specified orientation. For each subject, 180 sEMG signals were collected (12 gestures × 5 trials × 3 forearm orientations). Therefore, the proposed dataset comprises a total of 3420 sEMG signals (19 subjects × 180), as shown in Table I. To avoid muscle fatigue, 30 s time break was given between two successive trials and at least 10 to 15 minutes between two successive orientations.

*C. Data Acquisition*

In this study, a specific experimental setup was created for the sEMG signal acquisition as illustrated in Fig. 1(a) [33]. The device was configured with eight channels that facilitate the recording of the sEMG signal of the forearm muscle activity of a single subject.

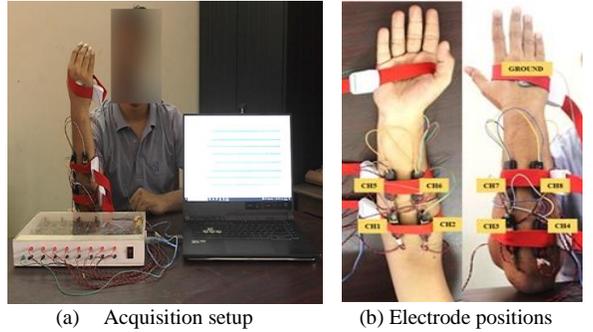
(a) Acquisition setup     (b) Electrode positions

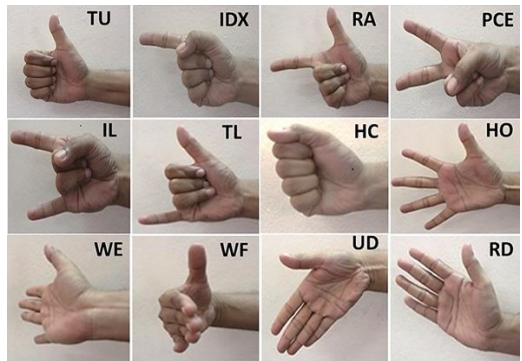
(c) Illustration of hand gestures

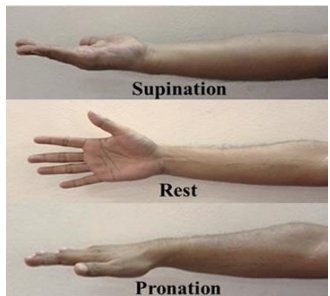
(d) Forearm orientations

**Fig. 1.** The sEMG signal recording setup and hand gestures with multiple forearm orientations.

The sampling frequency of each channel was 985 Hz at a resolution of 10 bits, ensuring precise capture of the sEMG signals at a high rate. The signal-to-noise ratio (SNR) value was up to 35 dB, depending on the selected gestures and the level of muscle contraction force. Among eight active electrodes of the system, four electrodes (modified from the MFI bar electrode, MFI Medical Equipment, Inc., USA) were denoted as CH1 to CH4 were placed circumferentially around the elbow and another four denoted CH5 to CH8 around the middle of the forearm as shown in Fig. 1(b). This arrangement helped capture a broad range of muscle signals from both the elbow and mid-forearm while performing the gestures. Additionally, the common ground electrode (MFI Medical Equipment, Inc., USA) was positioned on the hand's dorsal side as illustrated in Fig. 1(b).

Distal to the elbow joint, two electrodes (CH1, CH2) were positioned on the flexor digitorum superficialis and flexor carpi radialis of the anterior compartment of the forearm. On the posterior compartment of the forearm, CH3 and CH4 were positioned over the extensor digitorum and flexor carpi ulnaris. At the mid-forearm level, CH5 and CH6 were positioned in alignment with CH1 and CH2, respectively, targeting the same muscles on the forearm's anterior surface. Similarly, CH7 and CH8 were aligned with CH3 and CH4, focusing on the same muscles on the posterior side of the forearm. A sample of raw sEMG signals in the time and frequency domain of the RA gesture of a representative is shown in Fig. 2(a) and Fig. 2(b) respectively.

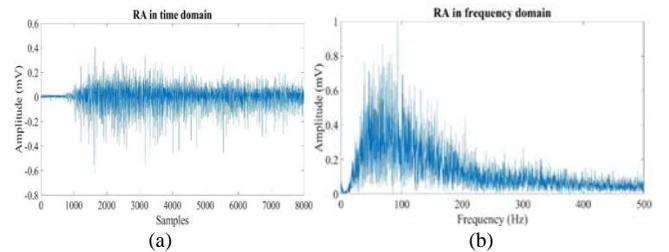
(a)     (b)
**Fig. 2.** Channel 1 raw sEMG signal of RA gesture in (a) time domain (b) frequency domain.

The format and content of the released database are described in Table II.

TABLE II
FORMAT AND CONTENT OF THE PROPOSED SEMG DATASET

| Folder | Sub-folder | Files | Description |
|---|---|---|---|
| Subject$_i$ | Pronation | gesture_name-j.mat | $i \in \{1,2,...,19\}$ represent subject index |
| | Rest | | $j \in \{1,2,...5\}$ represent the trial index |
| | Supination | | .mat file contains $8000 \times 8$ matrices indicating eight channels sEMG data of 8000 samples |
| Database Info | Readme | readme.txt | General information on dataset and explanation about how to read data files |
| | Subject and gestures info | Subject Details.csv, gestures_name.txt, gestures_sequence.jpg | Subject Details.csv contains subjects' information. Gestures_name.txt contains gestures list and gestures_sequence.jpg represents performed gestures with reading sequence. |
| | Acquisition setup info | Signal Acquisition.pdf | Signal Acquisition.pdf contains information about the device and electrode info. also contain electrode positioning info on the site of the elbow and the middle of the forearm position. |

The sEMG data were recorded using MATLAB software in the file format .mat, allowing users a convenient approach to access the sEMG recordings in a means of physical unit. Therefore, additional conversion can be avoided before signal preprocessing. The database contains nineteen folders corresponding to nineteen subjects. Each subject folder contains three subfolders that correspond to three forearm orientations (pronation, rest, supination). The raw data files of all hand gestures of all trials are in the subfolder named with considered forearm orientation. The naming convention of raw sEMG data of specific gestures is "gesture_name-trial.mat". For example, Hand_Close-4.mat contains the data for the HC gesture performed in the fourth trial. The database also contains a folder with add-on files providing supporting data about the participant's physical measurements, electrode attachment instructions, machine and electrode configuration, and gesture sequence with their illustrations. The dataset is publicly available on this website.

*D. sEMG Signal Quality Assessment*

Within this study, the quality and noise levels of the recorded sEMG signals were investigated using two standard signal quality parameters: Signal-to-Noise Ratio (SNR) and Signal-to-Motion artifact Ratio (SMR). SNR measures the amplitude of the sEMG signal relative to the background noise. It quantifies the difference between the muscle activity signal and the undesired electrical signal recorded at rest. A higher SNR value indicates relatively clean and more reliable signal, as it reflects a larger difference between the sEMG signal during muscle contraction and the baseline noise. This contributes to improving the robustness of sEMG pattern recognition systems. SNR was measured in decibels (dB) and calculated as shown in Equation (1). Afterward, the SNR was averaged across all channels, forearm orientations, trials, and rest periods. SMR quantifies the ratio of the sEMG signal amplitude (within the frequency range of 20 Hz to 500 Hz) to the motion artifact present in the raw signal. The motion artifact is caused by electrode displacement over the skin during muscle movement. It is hypothesized that the power spectral density (PSD) of the sEMG signal follows a linear pattern from 0 Hz to 20 Hz. A higher SMR value indicates that the sEMG signal contains less motion artifact, reflecting a higher-quality signal. SMR was calculated by dividing the overall power of the filtered signal by the motion artifact, as described in Equation (2).

$$SNR = 20 \log_{10} \frac{\sqrt{Raw\ EMG\ Signal_{RMS}^2 - Noise_{RMS}^2}}{Noise_{RMS}} \quad (1)$$

$$SMR = 10 \log \left[ \frac{\sum_{f=20}^{500} PSD}{\sum_{f=0}^{20} PSD} \right] \quad (2)$$

*E. Feature Extraction*

The effectiveness of extracted features on the sEMG pattern recognition system was investigated using five well-known feature extraction methods: signal normalized time domain features (SNTDF) [38], time domain descriptors (TDD) [39], temporal-spatial descriptors (TSD) [40], time-fusion-based time-domain descriptors (FTDD) [31], six-order autoregressive coefficients with root mean square value (AR-RMS) [41]. The SNTDF includes seven features: nonlinear mean absolute value, zero-, second-, fourth-, and sixth-order power spectra, first- and second-order average amplitude change, and correlation coefficients for each pair of channels. In this research, the SNTDF generated 84-dimensional feature space. The TDD includes five time-series features based on the sum of squared differences and properties of the Fourier transform. TDD produced a 40-dimensional feature space. The TSD includes zero-, second-, and fourth-order moments, sparseness, irregularity factor, Teager–Kaiser energy operator, coefficient of variation, and each pair of the difference of channels. TSD provided 252-dimensional feature space. The FTDD includes root squared zero, fourth and eighth order moments, sparseness, irregularity factor, waveform length ratio based on the original features, and nonlinearly transformed features. The FTDD produced 48 dimensions of feature spaces. The AR-RMS includes six-order autoregressive coefficients with root mean square value. It provided 56-dimensional feature space.

*F. Classification*

In this study, hand gesture recognition performance with three forearm orientations using the proposed dataset was evaluated using MATLAB 2020a (MathWorks, USA). Initially, the active sEMG signal was extracted in segments for a duration from 5 to 7 seconds out of the total 8 seconds of recorded signal. This segmentation of active signal depended upon the hand gesture activation from rest. Then, the raw EMG signal was preprocessed employing a bandpass filter of 20Hz to 450Hz with a notch filter at 50Hz to minimize high-frequency noise, movement-induced noise, and electrical interference. After that, the EMG signal was segmented with a disjoint windowing of 250 ms. Therefore, the total processing time lies within the recommended delay of 300 ms. Five widely acknowledged feature extraction techniques were employed to extract the features of active sEMG signal as mentioned in Section II(E). All the feature extraction methods created high-dimensional feature space. Therefore, the high dimensionality of the feature space of each feature extraction method was reduced to eleven (total hand gestures - 1) employing spectral regression discriminant analysis (SRDA). SRDA is a technique used to reduce the dimension particularly useful in the fields of pattern recognition as well as for classification problems [41], [42].

A 5-fold cross-validation method was employed to reduce the issue of data overfitting. In this case, all of the known and unknown orientation samples were split into training and testing samples. Then, the model was trained with the sample of known orientation only but tested with known and unknown orientation. Several state-of-the-art classification techniques have been investigated in the literature of sEMG-based gesture pattern recognition. One classification method is linear discriminant analysis (LDA) which is mostly well-known for its easy implementation, and efficient computation. Moreover, the performance is comparable to a more complicated algorithm. However, there are support vector machines (SVM) and K-nearest neighbors (KNN) which are considered as best-performing classification algorithms in sEMG-PR. The hyperparameters of the classifiers were fine-tuned using the '*Classification Learner*' app [43], [44]. Therefore, in this study, KNN with Euclidean distance and ten neighbors, SVM with Gaussian radial basis optimization (kernel scale was set to 3), and LDA were employed. Moreover, the classification performance of forearm orientation invariant hand gesture recognition has also been evaluated using robust deep learning (DL) -based models. The models included a 1D Convolutional Neural Network (1D CNN), Long Short-Term Memory (LSTM), Recurrent Neural Network (RNN), a combination of CNN and RNN, and a hybrid CNN with Bidirectional LSTM (BiLSTM). The architecture of the 1D CNN is implemented with the following parameters. The filter size of each Conv1D layer was 16 and the kernel size was 10. Rectified linear units (ReLU) were used as activation functions of Conv1D and dense layers. To prevent overfitting, dropout layers with rates of 50% and 30% were also added to the network with a learning rate of 1e$^{-5}$. The hybrid model CNN with BiLSTM was implemented with three Conv1D layers with progressively increasing filter sizes (64, 128, and 256 filters) and a kernel size of 7, 5, and 3 respectively. The BiLSTM layer with 512 units is employed with a dropout of 30% and a recurrent dropout of 20%. LeakyReLU activations and substantial dropout at 40% were added to prevent overfitting and the learning rate was 1e$^{-5}$. The LSTM model includes an LSTM layer with 128 units. The RNN-based model consists of a SimpleRNN layer with 16 units with a ReLU activation function. The hybrid CNN-RNN includes a Conv1D layer with a filter size of 16, kernel size of 10, and SimpleRNN layer with 16 units. For feature engineering, the SNTDF feature extraction method was employed for all DL-based models. The models were trained using 80% of the hand gesture data in the resting orientation of all subjects. Validation was performed using 10% of the data from each orientation: rest, supination, and pronation of all subjects. Finally, the trained models were tested with 10% of the data from each of these orientations of all subjects as well. For the performance evaluation of the classifiers, six statistical performance evaluation metrics including accuracy, sensitivity, specificity, precision, F1 score, and Matthew's correlation coefficient (MCC) were assessed [26].

$$Accuracy = \frac{TP+TN}{TP+TN+FP+FN} \quad (3)$$

$$Sensitivity = \frac{TP}{TP+FN} \quad (4)$$

$$Specificity = \frac{TN}{TN+FP} \quad (5)$$

$$Precision = \frac{TP}{TP+FP} \quad (6)$$

$$F1\ Score = \frac{2 \times Precision \times Sensitivity}{Precision+Sensitivity} \quad (7)$$

$$MCC = \frac{TN \times TP - FN \times FP}{\sqrt{(TP+FP)(TP+FN)(TN+FP)(TN+FN)}} \quad (8)$$

Hence, *TP*, *TN*, *FP*, and *FN* refer to true-positive, true-negative, false-positive, and false-negative gestures, respectively.

*G. Statistical Analysis*

To evaluate statistical differences, a two-way and three-way analysis of variance (ANOVA) with the Bonferroni correction test was performed to compare the F1 score across forearm orientations, feature extraction methods, and classifiers. In these statistical tests, a 5% significance threshold was considered. The *p*-value below 0.05 indicated the statistical significance of the F1 scores. In addition, only ANOVA was also performed to explore the impact of the window sizes on the F1 score of the best classifier.

III. RESULTS

*A. sEMG Signal Quality*

In this article, sEMG signals were recorded from the elbow to the middle of the forearm of 19 subjects while performing 12 active daily living hand gestures with 3 forearm orientations. Fig. 3 illustrates a sample signal of a representative participant, recorded from eight channels corresponding to all finger and wrist gestures. Two signal quality matrices, SNR and SMR were utilized to investigate the quality of the recorded sEMG signal. The SNR value was evaluated by subtracting the RMS value of the first 500 samples of the raw signal at rest from the RMS value of the raw sEMG signal for all gestures, as shown in (1). For each hand gesture, the SNR mean and standard deviation were evaluated across all subjects, channels, trials, and forearm orientations. SNR of all gestures across all channels considering three orientations are presented in Fig. 4. According to the experiment's findings, the SNR of all gestures ranges from 28.31 to 41.57 dB. Among the individual hand gestures, the highest SNR was achieved for HC at 41.57 dB, while HO generated a comparatively lower SNR at 33.85 dB. For the two single-digit gestures, TU achieved the highest SNR of 36.38 dB, whereas IDX generated an SNR of 31.3 dB. The SNR value of multi-digit gestures (RA, PCE, IL, TL) varied between 28.31 to 37.95 dB. However, TL achieved the highest SNR of 37.95 dB. Considering the group of wrist gestures (WE, WF, UD, RD), WE generated an SNR of 39.59 dB which is way more than the SNR of 31.28 dB for WF. The SNR of UD and RD were comparable, which were at 36.93 dB and 34.04 dB, respectively.

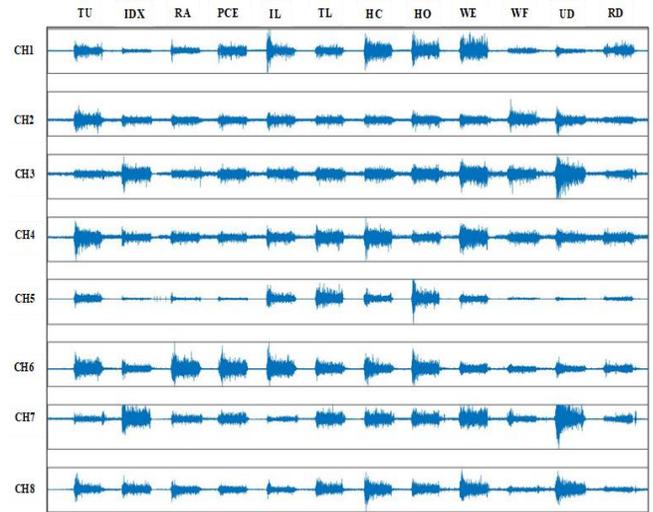

**Fig. 3.** Eight channel sEMG signals of different hand gestures.

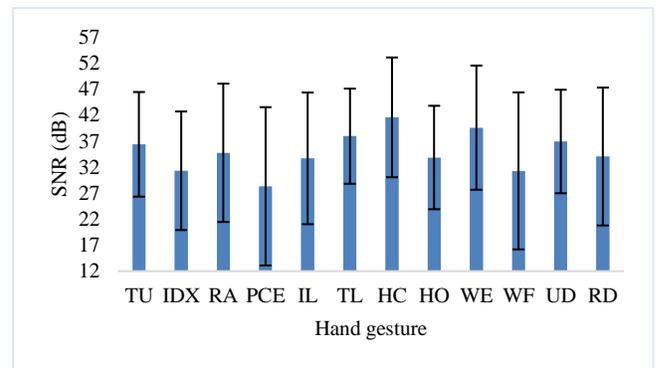

**Fig. 4.** The SNR values in dB of individual hand gestures under three forearm orientations across all channels; the bar represents the standard deviation of SNR per gesture.

The SMR of the sEMG signals recorded from the elbow to the middle of the forearm for all gestures varied between 7.63 and 9.96. The SMR values for all gestures under different orientations are illustrated in Fig. 5.

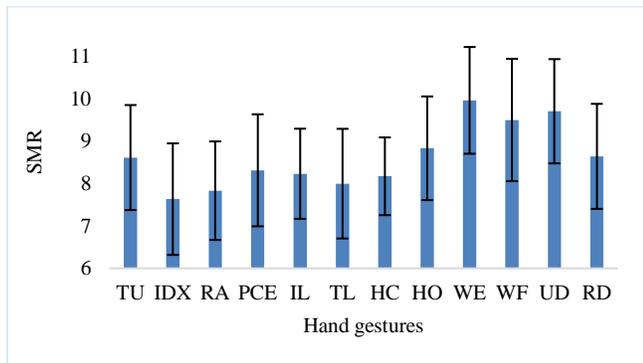

**Fig. 5.** The SMR values considering different forearm orientations of individual hand gestures.

The highest SMR value was achieved for the group of wrist gestures, with WE at 9.96, WF at 9.5, and UD at 9.7, while RD had a slightly lower SMR value of 8.64. Among the finger gestures, TU, PCE, and IL achieved SMR values of 8.61, 8.31, and 8.23, respectively. The rest of the finger gestures, IDX (7.63), RA (7.83), and TL (7.99) achieved comparatively lower SMR values. The SMR values of HO and HC gestures were comparable, at 8.83 and 8.17 respectively.

### B. Classification Performance

*Machine learning – based approach*

Hand gesture recognition performance under three forearm orientations was evaluated using five feature extraction methods and three state-of-the-art classifiers. All three classifiers were trained with hand gestures at rest orientation but tested with data across all orientations. The mean values of the performance evaluation metrics, along with the standard deviations, are presented in Table III. However, a brief overview of these results is graphically presented in Fig. 6 using the F1 score, as it represents true positive hand gesture recognition performance more precisely. In addition, a two-way ANOVA with Bonferroni correction was employed to evaluate the significance of the F1 scores for the selected classifiers and feature sets. The result of this study demonstrated that the hand gesture recognition performance under different forearm orientations was significantly dependent on the feature extraction methods ($p<0.05$) but independent of the classifiers ($p>>0.05$). Among the feature extraction methods, the best performance was achieved by the SNTDF feature ensemble, with an F1 score of 88.58%. The next best feature set was TDD, which produced an F1 score of 86.95%, while the AR-RMS

TABLE III
FOREARM ORIENTATION INVARIANT HAND GESTURE RECOGNITION PERFORMANCES WITH MULTIPLE CLASSIFIERS AND FEATURE EXTRACTION METHODS

|  |  | SNTDF | TSD | TDD | FTDD | AR-RMS |
|---|---|---|---|---|---|---|
| **Accuracy (%)** | LDA | 98.11±1.05 | 97.81±1.09 | 97.84±1.04 | 97.51±1.41 | 97.41±1.21 |
|  | SVM | 98.00±1.04 | 97.78±1.10 | 97.66±1.04 | 97.40±1.34 | 97.40±1.15 |
|  | KNN | 97.97±1.05 | 97.71±1.14 | 97.60±1.02 | 97.38±1.39 | 97.38±1.12 |
| **Sensitivity (%)** | LDA | 88.67±6.30 | 86.90±6.56 | 87.06±6.28 | 85.08±8.47 | 84.47±7.26 |
|  | SVM | 88.03±6.23 | 86.69±6.60 | 86.01±6.27 | 84.43±8.09 | 84.42±6.92 |
|  | KNN | 87.84±6.34 | 86.31±6.85 | 85.64±6.17 | 84.33±8.36 | 84.30±6.73 |
| **Specificity (%)** | LDA | 98.97±0.57 | 98.80±0.59 | 98.82±0.57 | 98.64±0.76 | 98.59±0.65 |
|  | SVM | 98.92±0.56 | 98.79±0.59 | 98.73±0.56 | 98.58±0.72 | 98.58±0.62 |
|  | KNN | 98.89±0.57 | 98.75±0.62 | 98.69±0.55 | 98.57±0.75 | 98.57±0.60 |
| **Precision (%)** | LDA | 90.03±5.29 | 88.62±5.44 | 88.54±5.47 | 87.28±6.93 | 86.44±5.80 |
|  | SVM | 89.44±5.52 | 88.28±5.62 | 87.96±4.95 | 86.64±6.50 | 86.30±5.32 |
|  | KNN | 89.28±5.59 | 88.11±5.64 | 87.42±5.24 | 86.63±6.56 | 86.10±5.38 |
| **F1 Score (%)** | LDA | 88.58±6.18 | 86.83±6.53 | 86.94±6.28 | 85.06±8.20 | 84.53±6.89 |
|  | SVM | 87.94±6.15 | 86.61±6.57 | 85.94±6.11 | 84.36±7.75 | 84.44±6.56 |
|  | KNN | 87.77±6.22 | 86.28±6.79 | 85.57±6.04 | 84.33±7.92 | 84.35±6.40 |
| **MCC** | LDA | 0.88±0.06 | 0.86±0.06 | 0.86±0.07 | 0.84±0.08 | 0.83±0.07 |
|  | SVM | 0.87±0.06 | 0.85±0.06 | 0.85±0.06 | 0.83±0.08 | 0.83±0.06 |
|  | KNN | 0.87±0.06 | 0.85±0.07 | 0.84±0.06 | 0.83±0.08 | 0.83±0.06 |

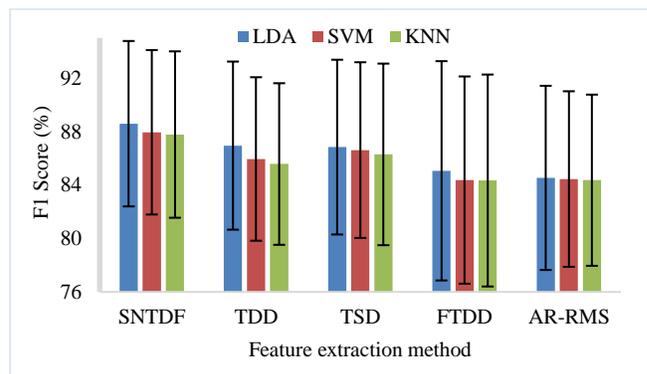

**Fig. 6.** The sEMG-PR performances with different forearm orientations and feature extraction methods using three classifiers.

feature set had the lowest F1 score of 84.53%.

The performance of the classification of hand gesture recognition was analyzed utilizing machine learning algorithms including LDA, KNN, and SVM along with five feature extraction methods. Among the three classification algorithms, the LDA using the SNTDF feature ensemble achieved the best performance, with an F1 score of 88.58% as illustrated in Fig. 6. The performance of SVM and KNN with the SNTDF feature ensemble was comparable, with an F1 score of 87.94% and 87.77%, respectively. The lowest performance was obtained by KNN with FTDD, producing an F1 score of 84.33%. Thus, the result indicates that the LDA classifier integrated with the SNTDF feature ensemble, outperforms the other two standard classification algorithms.

A confusion matrix was utilized to represent the classification performance visually and obtain an in-depth understanding of the result as shown in Fig. 7. In this case, the best-performing feature extraction method, SNTDF, with the LDA classification algorithm was used to evaluate the confusion matrix across all subjects. The model was trained with rest, however tested with three orientations. In the confusion matrix, the recall (or sensitivity) of the respective gesture is represented, reflecting the classifier's performance in terms of the true positive rate. The confusion matrix represents the true positive rate for each gesture along the diagonal. Hence, compared to single-digit gestures, the true positive rate for multi-digit and wrist gestures was more satisfactory. The most accurately labeled class is the TL gesture, while the IDX and TU gestures had higher misclassification rates. Despite this, the average recall and F1 scores across all subjects were 88.67% and 88.58%, respectively.

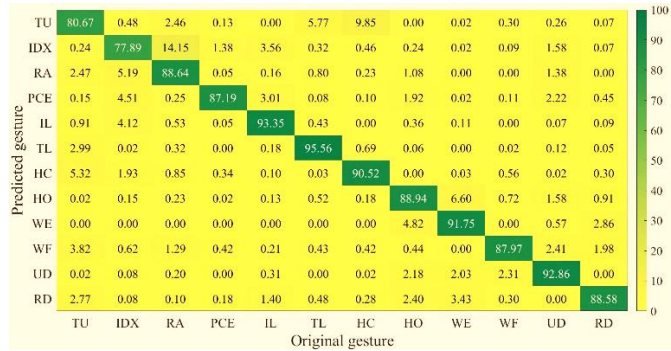

**Fig. 7.** Confusion matrix illustrating the recall across all subjects using SNTDF with LDA; trained with rest but tested with all orientations.

In this study, the performance of hand gesture recognition across three forearm orientations was further analyzed by consisting only of finger gestures (both single and multi-digit) and the other consisting only of wrist gestures. The same set of classifiers and feature extraction methods was applied to both groups. The results showed that for the finger gestures group, LDA with SNTDF achieved the highest F1 score of 89.40%, outperforming other combinations of classifiers and feature extraction methods, as shown in Fig. 8(a). Additionally, the F1 score of KNN was comparable to SVM with all five feature ensembles. In contrast, the experimental result for the wrist gestures group showed that the F1 score of all three classifiers and feature sets were closely comparable, as illustrated in Fig. 8(b). Notably, LDA with AR-RMS achieved the highest F1 score of 93.02%, outperforming all combinations of feature sets and classifiers for wrist gestures.

In this work, the classification performance of hand gesture recognition was analyzed for two forearm locations: the elbow and mid-forearm. Channels CH1-CH4 were placed near the elbow, while CH5-CH8 were positioned circumferentially on the mid-forearm. The classification performance of CH1-CH4 (elbow), CH5-CH8 (mid-forearm), and CH1-CH8 (all channels) was analyzed using the LDA classifier with five feature extraction methods. The classifier was trained using data of the rest orientation and tested with data of all orientations. The performance results for the two locations and all channels are graphically represented in Fig. 9. The results indicate that the mid-forearm location achieved a higher F1 score of 86.71% compared to the elbow, which attained an F1 score of 82.4% when using the SNTDF method. Additionally, utilizing all channels, the F1 score increased to 88.58% with SNTDF, surpassing the F1 scores of individual locations. A two-way ANOVA with Bonferroni correction was conducted to evaluate the significance of the F1 scores for the three cases: elbow vs. mid-forearm, elbow vs. all channels, and mid-forearm vs. all channels, as well as for the feature extraction methods. The analysis revealed that the performance of hand gesture recognition was significantly dependent on the channel location and the feature extraction methods ($p < 0.05$).

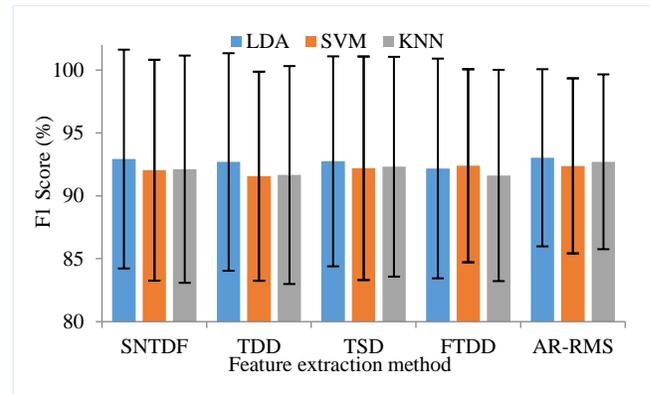

(a)

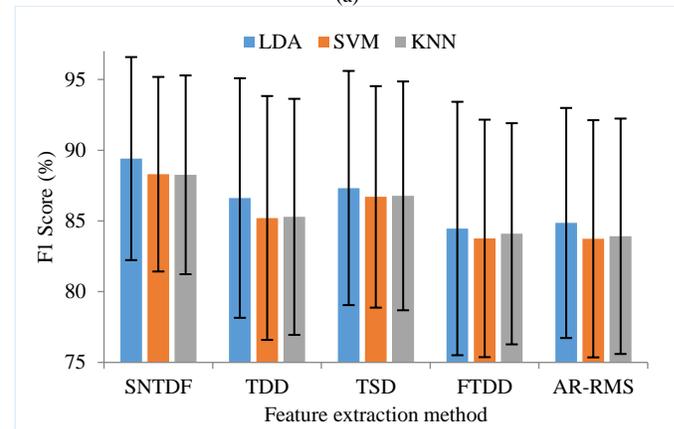

(b)

**Fig. 8.** The sEMG-PR performances with different forearm orientations and feature extraction methods using three classifiers. (a) finger gestures (b) wrist gestures.

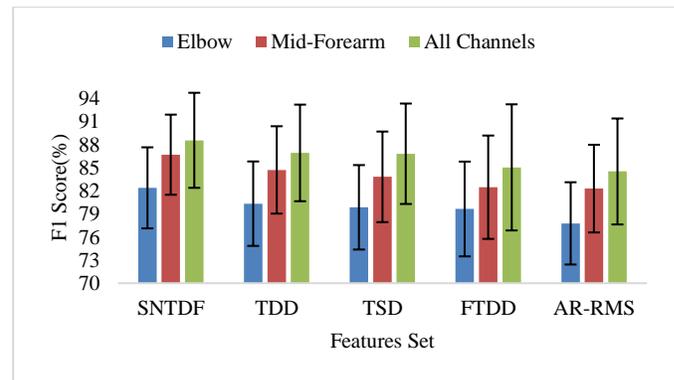

**Fig. 9.** The sEMG-PR performances considering different electrode/channel locations with three forearm orientations and feature extraction methods using the LDA classifier.

The hand gesture recognition performance with different forearm orientations was also evaluated across variable window sizes. In this case, the best-performing classifier LDA with SNTDF was used for performance evaluation. The considered window sizes were 50 ms, 100 ms, 150 ms, 200 ms, 250 ms, 300 ms, and 350 ms. It was observed that the performance, measured by the F1 score, rose progressively as the span of window sizes increased. When the window length was 100 ms, the F1 score was 87.04%; however, when the window size was 300 ms, it improved to 89.01%. The performance of the LDA classifier with SNTDF with various window sizes is illustrated in Fig. 10. To find statistical significance, only ANOVA was conducted employing variable window sizes. Results showed that performance significantly increased up to 100 ms with an F1 score of 87.04%. However, beyond 100 ms, there was no significant increase in F1 score up to 300 ms window size.

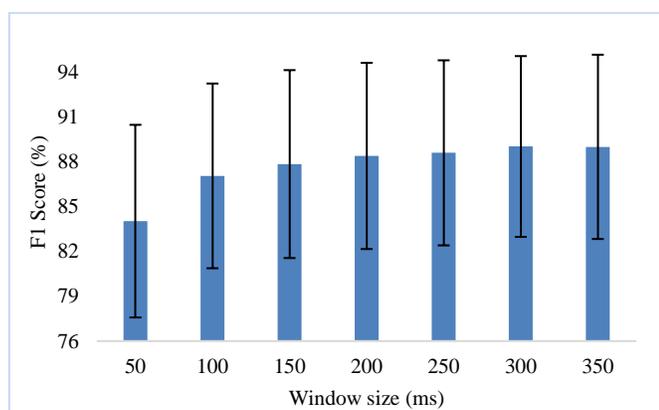

**Fig. 10.** The forearm orientation invariant hand gestures recognition performance with variable window sizes using LDA with SNTDF.

*Deep Learning-based approach*

In this study, several DL-based models were utilized to evaluate their effectiveness in forearm orientation-invariant hand gesture classification using the proposed dataset. The models included a 1D CNN, LSTM, RNN, a combination of CNN and RNN, and a hybrid CNN with BiLSTM. Table IV outlines the classification performance of these DL models. The experimental results indicate that the 1D CNN achieved the highest performance, with accuracy, sensitivity, specificity, precision, F1 score, and MCC of 85.28%, 85.77%, 98.65%, 84.54%, 85.01%, and 0.83, respectively. The second-best performance was achieved by the LSTM model, which achieved an accuracy of 82.53%, sensitivity of 83.11%, specificity of 98.41%, precision of 81.80%, an F1 score of 82.24%, and an MCC of 0.81. In contrast, the CNN-RNN model exhibited the lowest performance, with accuracy, sensitivity, specificity, precision, F1 score, and MCC of 70.64%, 71.20%, 97.33%, 69.44%, 69.83%, and 0.67, respectively.

### C. Performance Comparison

In this article, we compared and validated the EMG-PR performance of hand gesture recognition using the proposed dataset against existing work, as shown in Table V. Khushaba *et al.* [31] investigated the combined influence of forearm orientations and muscular contraction force levels (low, medium, high) on the sEMG signal. They employed the TSD feature extraction method with SVM to classify six hand and wrist gestures. When they trained the classifier with the data of six hand gestures from rest orientation and tested with pronation and supination, a performance of 60.67% was achieved for intact-limbed subjects. However, the performance improved to 76.77% when the classifier was trained with rest data from all force levels and tested with all orientations and force levels for amputees only. Rajapriya *et al.* [37] proposed a wavelet bispectrum-based features set to evaluate EMG-PR performance, considering the combined effect of forearm orientations and contraction force levels. They achieved a performance of 86.43% when the classifier was trained with rest data from all force levels and tested with all three orientations across all force levels. Additionally, Islam *et al.* [26] achieved forearm orientation and muscle force-invariant performance of 78.44% when the classifier was trained with rest data only but tested with all orientations.

Table IV
FOREARM ORIENTATION INVARIANT HAND GESTURE RECOGNITION PERFORMANCES WITH DL-BASED MODELS

| DL Model | Accuracy (%) | Sensitivity (%) | Specificity (%) | Precision (%) | F1-Score (%) | MCC |
|---|---|---|---|---|---|---|
| 1D CNN | 85.28 | 85.77 | 98.65 | 84.54 | 85.01 | 0.83 |
| LSTM | 82.53 | 83.11 | 98.41 | 81.80 | 82.24 | 0.81 |
| RNN | 77.60 | 78.63 | 97.97 | 76.80 | 77.00 | 0.75 |
| CNN-BiLSTM | 82.15 | 82.69 | 98.38 | 81.40 | 81.76 | 0.80 |
| CNN-RNN | 70.64 | 71.20 | 97.33 | 69.44 | 69.83 | 0.67 |

In this study, the forearm orientation-invariant sEMG-PR performance improved to 88.58% using our proposed dataset, as shown in Table V. The classifier was trained with hand gestures in the rest orientation and tested across all orientations. Compared to related datasets considering forearm orientations, the proposed multichannel sEMG dataset, which includes a larger number of subjects and hand gestures, demonstrated significantly improved EMG-PR performance, highlighting the novelty of this dataset.

## IV. DISCUSSION

EMG-PR is an efficient method for decoding muscular activity during physical movements and regaining the lost functionality for amputees. However, several factors can alter the time and frequency domain properties of the EMG signal, resulting in the degradation of EMG-PR performance. Forearm orientation is a crucial factor that limits the hand gesture recognition performance of EMG-PR systems [37]. The forearm muscles pronator teres and pronator quadratus are primarily responsible for pronation, while the supinator performs supination [45]. The inactive state of these three muscles is considered the rest orientation. For daily activities,

TABLE V

THE FOREARM ORIENTATION INVARIANT HAND GESTURE RECOGNITION PERFORMANCE COMPARISON

| Reference | Dynamic Factor | Number of Subjects | Number of Gestures | Number of Trials | Feature Extraction Method | Classification | | Classifier | EMG-PR Performance (%) |
|---|---|---|---|---|---|---|---|---|---|
| | | | | | | Training | Testing | | |
| Khushaba et al. [31] | 3 forearm orientations with 3 muscle fore levels (low, medium, and high) | 13 (12 healthy & 1 amputee) | 6 | 3 | TD-PSD and 3-D accelerometer features | Rest orientation | Pronation and supination | SVM | 60.67 (healthy) |
| | | | | | | Rest orientation from all force levels | All orientations from all force levels | SVM | 76.77 (amputee) |
| Rajapriya et al. [37] | 3 forearm orientations with 3 muscle fore levels (low, medium, and high) | 10 | 6 | 3 | Wavelet bispectrum-based features set | Rest orientation from all force levels | All orientations from all force levels | LDA | 86.43 |
| Islam et al.[26] | 3 forearm orientations with 3 muscle fore levels (low, medium, and high) | 10 | 6 | 3 | SNTDF, 15-order autoregression coefficients, and accelerometer features | Rest orientation from a medium force level | All orientations from all force levels | KNN | 78.44 |
| Proposed Work | 3 forearm orientations | 19 | 12 | 5 | SNTDF | Rest orientation | All orientations | LDA | 88.58 |
| | | | | | | | | SVM | 87.94 |

various hand gestures are performed with different forearm orientations to accomplish useful tasks. When a hand gesture with a particular forearm orientation is performed, the muscles responsible for the associated orientation and the accompanying gesture are both recruited. As a result, forearm orientation changes the sEMG signal of the corresponding hand gesture by overlaying the sEMG signal from muscles responsible for forearm orientation [46]. Consequently, forearm orientation significantly degrades the performance of gesture classification systems [47], [48], [49]. A few studies have investigated the impact of forearm orientation variation on sEMG signals [26], [31], [37]. Despite these efforts, publicly available sEMG datasets that incorporate forearm orientation remain limited, narrowing the scope of research to improve forearm orientation-invariant EMG-PR systems.

This study proposes a novel sEMG dataset recorded from eight MFI EMG electrodes positioned on the elbow and the middle of the forearm, with participants performing twelve hand gestures in three forearm orientations. The sEMG signal quality of this dataset was validated in terms of SNR and SMR. Experimental results demonstrated that the SNR of all gestures ranged from 28.31 dB to 41.57 dB, indicating high-quality sEMG signals with minimal noise artifacts. Among all gestures, the HC gesture achieved the highest signal strength at 41.57 dB, while the PCE gesture produced the lowest signal strength at 28.31 dB. This difference may be due to the number of muscles activated for different hand motions. For the HC gesture, multiple forearm and hand muscles are activated, resulting in stronger signal strength. Conversely, the PCE gesture primarily recruits the extensor digitorum and extensor indicis, leading to comparatively lower signal strength. In terms of other signal quality metrics, SMR values ranged from 7.63 to 9.96, indicating that the EMG signals were minimally affected by movement-related noise and electrical interference. Thus, the high SNR and SMR values confirm the high quality and signal strength of the proposed dataset.

The dataset was further evaluated for forearm orientation-invariant hand gesture recognition performance using five feature extraction methods and three machine learning classifiers. Data for twelve hand gestures in the rest orientation were used to train the classifiers, while gesture data from all orientations were used for testing. The SNTDF feature extraction method consistently outperformed others across all classifiers, likely due to its ability to enhance force-invariant pattern recognition through higher-order indirect frequency information, nonlinear signal transformations, and correlation coefficients [38]. This method maintains consistency across varying amplitude levels due to muscle contraction forces. A three-way ANOVA indicated that recognition performance was significantly dependent on forearm orientations ($p < 0.001$) and feature extraction methods ($p < 0.001$), but independent of classifiers ($p \gg 0.05$). The classification performance was further illustrated using a confusion matrix, which showed higher misclassification rates for single-digit gestures such as IDX and TU, whereas multi-digit and wrist gestures were more accurately detected. This higher error rate for single-digit gestures may be due to the lower muscle activation required compared to more complex gestures or variations in forearm orientation with varying contraction strengths. Training models with data from a broader range of real-life forearm orientations could help mitigate this issue.

Additionally, the dataset's performance was analyzed by dividing gestures into two groups: finger gestures and wrist gestures. The results showed a slight increase in performance for the finger gesture group compared to using all gestures, while the wrist gesture group showed a 4.34% increase in performance with the SNTDF feature method and LDA classifier. The performance of forearm orientation invariant sEMG-PR was analyzed for two forearm locations: elbow and mid-forearm. The experimental result revealed improved performance for the mid-forearm location compared to the elbow, with performance being significantly dependent on the location and the features. The impact of window size on sEMG-PR performance was also evaluated, with sizes ranging from 50 ms to 350 ms in 50 ms intervals [50]. A one-way ANOVA test found that a 50 ms window size was significant, while other sizes were not. The performance increased from 84% to 89% as the window size ranged from 100 ms to 350 ms, indicating that pattern recognition performance can be consistently improved with varying window sizes, enhancing system robustness. Regarding the performance of five DL-based models utilizing the proposed dataset, the 1D CNN achieved the highest F1 score of 85.01%, outperforming the other DL models. This can be attributed to CNN's ability to effectively capture local temporal patterns inherent in sequential data.

The comparison of forearm orientation-invariant sEMG-PR classification performance using the proposed dataset demonstrates a significant improvement over existing studies on the impact of forearm orientations on sEMG, as shown in Table V. To our knowledge, the Rami-Khushaba dataset [31], which included forearm orientations, achieved a performance of 60.67% with SVM when trained with rest data and tested with data from pronation and supination. In contrast, our dataset demonstrated a significantly improved performance of 87.94% with the same classifier when trained with rest data and tested across all orientations. The highest classification performance with the proposed dataset was achieved using the SNTDF feature set and the LDA classifier, yielding an F1 score of 88.58%, representing a 2.15% improvement over the results of Rajapriya *et al* [37]. Moreover, previous studies included forearm orientations with varying muscular contraction levels, this study improved performance by training with rest orientation data only and testing across all orientations using the proposed dataset.

All the above evaluations were carried out to demonstrate the high signal quality and validation of the proposed dataset. The high competence results showcase the dataset's robustness, reliability with high signal strength, and improved performance of forearm orientation invariant EMG-PR. The dataset also comprises sEMG signals from many subjects and gestures compared to related studies on forearm orientations, making it suitable as a benchmark for further development or for enhancing the quality of hand gesture recognition systems. Concerning future research directions, the dataset could be enhanced by considering more dynamic gestures such as pulling, pushing, grasping, and writing. Electrode displacement is a crucial factor in limiting the myoelectric pattern recognition system performance [52]. Addressing this factor along with forearm orientation may contribute to greater enhancement of the classification performance of the sEMG pattern recognition system. In addition, a larger population can be included, encompassing both subjects with intact-limbed and amputees [51]. This factor would increase its robustness and applicability. Finally, the performance of deep learning-based models using this dataset can be further enhanced by incorporating dropout or batch normalization layers and optimizing hyperparameters.

## V. Conclusion

In this research, we have introduced a novel sEMG dataset obtained from the forearm of nineteen healthy Bangladeshi subjects. The dataset includes recordings of twelve daily living hand gestures, performed with three forearm orientations (pronation, rest, supination) and captured from two electrode locations (elbow to mid-forearm). According to our findings, it is the first dataset of its kind focused on forearm orientations and recordings from two locations of the forearm. The dataset was validated against the existing sEMG datasets and analyzed using signal quality metrics (SNR and SMR). In addition, forearm orientation invariant hand gesture classification is performed considering feature extraction methods, variable window size, and popular machine learning and deep-learning algorithms. The extensive analysis of the proposed dataset, resulting in high signal strength and improved performance of the forearm orientation invariant hand gesture recognition system implies the efficiency, reliability, and robustness of the proposed forearm orientation invariant sEMG dataset. Therefore, the proposed sEMG dataset across multiple forearm orientations can serve as a standard basis for developing gesture-based recognition systems, conducting clinical research on sEMG, and advancing prosthetic devices and human-computer interface applications.

## Acknowledgment

The authors would like to express their deepest gratitude to the subjects for their voluntary contribution to this research.

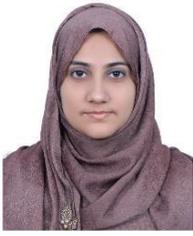
**Umme Rumman** received the B.Sc. and M.Sc. degrees in Computer Science and Engineering from the University of Rajshahi, Rajshahi, Bangladesh, in 2014 and 2015, respectively. She is an Assistant Professor at the Department of Computer Science and Engineering, Varendra University, Rajshahi. Her areas of research interest include human-machine interfacing, biomedical signal processing, bioinformatics, high-performance computing, and machine learning.

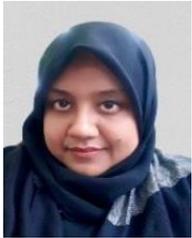
**Arifa Ferdousi** was born in Rajshahi, Bangladesh in 1984. She received her B.Sc. and M.Sc. degree in Information and Communication Engineering, from the University of Rajshahi, Bangladesh. Currently, she is a PhD fellow in the Electronic and System Engineering Lab, Department of Electrical, Electronic and System Engineering, National University of Malaysia, Malaysia. She served as a lecturer in the Department of Electronic and Telecommunication Engineering of Prime University, Dhaka Bangladesh, from 2009 to 2012. From 2012 she joined as a lecturer in the Department of Computer Science and Engineering, at Varendra University, Rajshahi, Bangladesh. Later in 2015, she was promoted to the designation of Assistant Professor. Her research interest includes sEMG pattern recognition and deep learning.

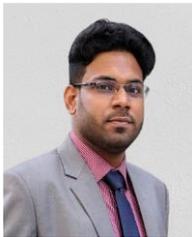
**Bipin Saha** is a Machine Learning Engineer at Business Automation Ltd., based in Rajshahi, Bangladesh. He obtained his B.Sc. in Electrical and Electronic Engineering from the University of Rajshahi in 2022. His research focuses on Vision-Language Modeling, Visual Perception, and the design of Intelligent Systems.

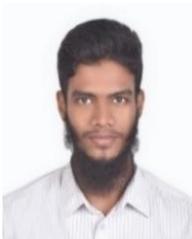
**Md. Sazzad Hossain** received a B.Sc. degree in electrical and electronic engineering from the University of Rajshahi, Rajshahi, Bangladesh, in 2021. His current research interests include electromyography (EMG), human-machine interfacing, and biomedical instrumentation.

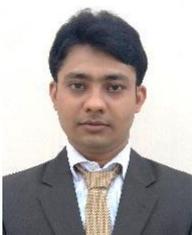
**Johirul Islam** was born in Meherpur, Bangladesh in 1989. He received his Ph.D. degree in 2022 from the Faculty of Engineering, University of Rajshahi, Bangladesh. He received his B.Sc. and M.Sc. degrees in Applied Physics and Electronic Engineering from the same university in 2011 and 2012, respectively. Currently, he is an Associate Professor in the Department of Physics, at Rajshahi University of Engineering and Technology, Bangladesh. His research interests include feature engineering, EMG pattern recognition, human-machine interfacing, biomedical instrumentation, and embedded systems.

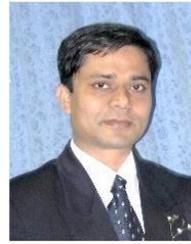
**Shamim Ahmad** received his Doctor of Engineering in Electrical Engineering from Chubu University, Japan. He got his B.Sc. and M.Sc. degrees in Applied Physics and Electronic Engineering from the University of Rajshahi, Rajshahi, Bangladesh. Following that he worked as a post-graduate research student in the Department of Computer Engineering, Inha University, South Korea. Currently, he is working as a Professor in the Department of Computer Science and Engineering of the University of Rajshahi, Rajshahi, Bangladesh. His areas of research are bioinformatics, embedded systems, and image processing.

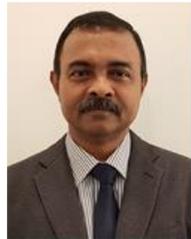
**Mamun B. I. Reaz,** PhD '07, is the Dean of the School of Engineering, Technology and Sciences at the Independent University, Bangladesh, and Professor in Electrical and Electronic Engineering. Previously, he was a Professor at the Universiti Kebangsaan Malaysia, Malaysia. His scientific specialization is in the areas of IC Design, Biomedical application IC, Biomedical sensors, and Smart Home. Mamun Bin Ibne Reaz has published more than 400 scientific articles and is a recipient of more than 70 research grants. Since 2020, he has been listed among the world's top 2% of scientists by Stanford University Data for "Updated science-wide author databases of standardized citation indicators". He was a Senior Associate of the Abdus Salam International Centre for Theoretical Physics (ICTP), Italy since 2008, and is presently, the Coordinator of the ICTP EAU Affiliated Centre in UKM, Malaysia. Mamun Bin Ibne Reaz has an undergraduate and graduate degree in Applied Physics and Electronics from the University of Rajshahi, Bangladesh, and a doctoral degree in VLSI Design from the Ibaraki University, Japan.

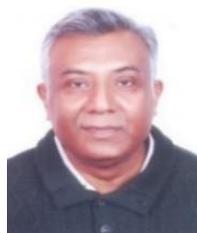
**Md. Rezaul Islam** received the B.Sc. and M.Sc. degrees in Applied Physics and Electronics from the University of Rajshahi, Bangladesh, in 1982 and 1983, respectively. He is a Professor in the Department of Electrical and Electronic Engineering, at the University of Rajshahi, Rajshahi, Bangladesh. His area of research includes electronics, biomedical instrumentation, biomedical signal processing, and embedded systems.